**Influence of different exchange-correlation potentials on twisted structures of bilayer XS$_2$ (X= Mo, Cr)**

Feng Sun[1], Ting Luo[1], Lin Li[2], Aijun Hong[1*], Cailei Yuan[1*], Wei Zhang[3]

[1]*Jiangxi Key Laboratory of Nanomaterials and Sensors, School of Physics, Communication and Electronics, Jiangxi Normal University, Nanchang 330022, China*

[2] *Material Technology Institute, Yibin University, Yibin 644000, China*

[3] *State Key Laboratory of Hydroscience and Engineering, Department of Energy and Power Engineering, Tsinghua University, Beijing 100084, China*

Correspondence and requests for materials should be addressed to A.J.H. (email: 6312886haj@163.com or haj@jxnu.edu.cn) and C.L.Y. (email: clyuan@jxnu.edu.cn).

**Abstract:**

In this work, we employ the LDA, GGA and GGA with four vdW corrections to study crystal and electronic structures of bilayer transition metal dichalcogenides (TMDs) with different twist angles. We find the GGA interlayer distance of bilayer MoS$_2$ has good agreement with experimental value while vdW correction method still needs to be further improved. Our results indicate the GGA interlayer distances of bilayer XS$_2$ (X= Mo, Cr) with twist angles are smaller than that of normal bilayer, which is the opposite in the LDA case. The GGA results show that reduced bandgap is due to the reduction of interlayer distance and, flat valley and conductivity bands appear owing to twist angle. Our study not only supports valuable information for application possibility of twisted two-dimensional (2D) materials but also stimulates more related research.

## I. Introduction

Since graphene was stripped from graphite, researchers have devoted a great deal of vigor to study 2D materials. Finding new 2D materials and modifying crystal and electronic structures of the existing are two hot spots of current researches. In addition to traditional methods, some new strategies such as strain, electric field, and twist angle are applied to the modification of 2D materials for obtaining novel physical properties. For example, softening phonon and band edge change such as band gap variation and a transition from direct to indirect band gap were detected in strained transition-metal dichalcogenide (TMD) 2D crystal. Theoretically, it is reported that vertical electric field can induce tunable bandgap of bilayer TMDs. Bilayer graphene with magic angle (namely Moiré superlattice) was predicted to can lead to strong coupling between layers, remarkably flat band and nearly zero of the Fermi velocity.

Crystal and electronic structures of TMDs with twist angle have been researched extensively also and exhibit interesting physical and chemical properties. However, interlayer distance dependency of twist angle is rarely studied. Previous studies have either used a fixed interlayer distance or adopt an exchange-correlation potentials functional to get a result that agrees with the experiment. As known, the interlayer distance has a non-negligible impact on electronic structure of TMDs although van der Waals (vdW) force between layers is weak. Previous report has shown that the increase of interlayer distance can enlarge the band gap size of normal bilayer $MoS_2$. Furthermore, more importantly, influences of different exchange-correlation potentials on twisted structures are often overlooked. Thus, it is necessary and of significance to systematically study it.

$XS_2$ (X= Mo, Cr) are as representative members of layered TMDs and have huge potential application prospects in energy and information fields. Especially, 2D $MoS_2$ has been extensively studied experimentally and theoretically, which can supports valuable information for other study on the other TMDs. Although pure phase of $CrS_2$ has not been prepared successfully, previous experimental study reported the monolayer $CrS_2$ with multiphase coexisting could be prepared via the chemical vapor deposition (CVD). On the other hand, theoretically, it was reported that the 2H structure is the ground state of CrS$_2$ monolayer and the formation energy of 2H CrS$_2$ monolayer is lower than that of most TMDs monolayers. These imply pure 2H-CrS$_2$ phase is the most likely to be synthesized first. Therefore, in this work, we choose two layered TMDs XS$_2$ (X= Mo, Cr) as research objects. We select six correlation-exchange functionals and adopt two optimization strategies to optimize the twisted structures with three twist commensurable angles. The calculation results show the GGA interlayer distance of bilayer MoS$_2$ has good agreement with experimental value while vdW correction methods still needs to be further improved. The GGA results indicate the interlayer distances of bilayer TMDs with twist angles are smaller than that of normal bilayer, which is the opposite in the LDA case. Moreover, the reduced GGA interlayer distance and twist angle can leads to the reduced band gap and flat valley and conductivity bands respectively.

## II. COMPUTATIONAL METHODS

We use the Vienna Ab Initio Simulation Package (VASP) to perform the first-principles calculations with the local density approximation (LDA) and the generalized gradient approximation (GGA), respectively. For GGA, the Perdew–Burke–Ernzerhof (PBE) exchange-correlation functionals with four van der Waals (vdW) corrections are employed. A slab of 15 Å thick is added to periodic structure for clearing away interaction between bilayers. We adopt two optimization strategies: the fully optimization and the fixed *c*-axis length optimization. The cutoff energy, total energy and force criterions and *k*-point mesh are set to 500 eV, 10$^{-5}$ eV, 0.05 eVÅ$^{-1}$ and 5×5×1 in all optimization calculations. Then, we adopt denser *k*-point mesh of 15×15×1 for accurate self-consistent calculation (SCF). For band structures calculations, the total number of *k*-point along the high-symmetry lines is 110.

## III. RESULTS AND DISCUSSIONS
**A crystal structure**

In order to obtain twisted structures of bilayer $XS_2$ (X= Mo, Cr), one can define commensurate cell vectors $A$ ($m$, $n$, $a_1$, $a_2$) and $B$ ($m$, $n$, $a_1$, $a_2$) :

$$A = na_1 + ma_2, \quad (1)$$

$$B = -ma_1 + (m + n)a_2, \quad (2)$$

where $m$ and $n$ are integers, and the basis vectors $a_1$ and $a_2$ with the lattice constants $a_0$ of TMDs are defined by:

$$a_1 = (\frac{\sqrt{3}}{2}, -\frac{1}{2})a_0, \quad (3)$$

$$a_2 = (\frac{\sqrt{3}}{2}, \frac{1}{2})a_0. \quad (4)$$

When one layer rotates by twist angle $\theta$, the vector $A$ on this layer coincides with $B$ vector on the other layer, and then twisted structures is formed. There is such a relationship between $\theta$ and ($m$, $n$):

$$\cos\theta = \frac{m^2 + 4mn + n^2}{2(m^2 + mn + n^2)}. \quad (5)$$

Thus, the total number of atoms in the in the primitive cell for $XS_2$ (X= Mo, Cr) twisted structures is

$$N = 6(m^2 + mn + n^2). \quad (6)$$

Considering the limited computing resources, we only study three representative twist angle structures (see Fig. 1) that correspond to $m$, $n$ and $N$ listed in Table I. It is worth mentioning that a 2×2×1 supercell containing 24 atoms is adopted for calculations of normal structure with $\theta = 0°$.

Table I twist angles $\theta$ corresponding to ($m$, $n$) and total number $N$ in the primitive cell.

| ($m$, $n$) | $\theta$ | $N$ |
|---|---|---|
| (1, 1) | 0° | 6 |
| (2, 3) | 13.2° | 114 |
| (1, 2) | 21.8° | 42 |
| (1, 3) | 32.2° | 78 |

The interlayer distance (ID) is defined by the distance between Mo atomic layers (see Fig. 1(a)). IDs of the bilayer $XS_2$ (X= Mo, Cr), attained by fully optimization strategy (marked by A) and the optimization method of fixed *c*-axis length (marked by B), are plotted in Fig. 2 and Fig. 3, respectively. For normal structure of bilayer $MoS_2$, the interlayer distance (ID) by A ranges from 6.0255 Å to 7.2334 Å, and that by B ranges from 6.4264 Å to 7.7293 Å that is agreement with previous work . No matter which optimization strategy, the maximal and minimal IDs come from GGA and LDA, respectively. It is quite acceptable since pure GGA and LDA normally overestimates and underestimates lattice constants, respectively. The experimental ID of bilayer $MoS_2$ is 7.0 Å . Obviously, the GGA ID (7.2334 Å) of bilayer by A strategy best matches the experimental data, although their difference reaches up to 3%. In order to validate the advantages of the GGA approach further, we take into account spin-orbit coupling (SOC) effects to perform optimization calculations and found that the interlayer distance of bilayer $MoS_2$ structure reduces to 7.0018 Å that ties in very well with the experimental data. The GGA ID of normal structure is larger than that of twist angle structures. The LDA ID of normal structure optimized by A strategy is smaller than that of twist angle structures, which is in agreement with previous work . However, the LDA IDs by B strategy hardly change with the twist angle. LDA IDs are sensitive to the optimization strategy. These may suggest that the GGA results have more credibility than the LDA results.

Interestingly, in the most cases, the GGA vdW correction leads to the opposite trend to pure GGA: twist angle causes larger ID. The GGA-DFT-D2 ID of zero angle structure by A strategy is smaller ~0.1 Å than that of twist angle structures, although corresponding IDs by B strategy are almost the same, which has the difference between each other is less than 0.01 Å. For the GGA-vdW-DF, the ID of 13.2° structure by A is smaller; nevertheless, it by B is larger than that of zero angle structure. The IDs of 21.8° and 32.2° structures by both A and B strategy are larger than that of zero angle structure. For the GGA-dDsC, the IDs of 13.2°, 21.8° and 32.2° structures by A strategy are larger about 0.25 Å than that of zero angle structure. However, their IDs by B strategy are smaller than that of zero angle structure. The case of GGA-vdW-DF2 is the same as the GGA-dDsC. These imply GGA-vdW method still needs to be further optimized at least in the case of prediction on interlayer distances of twisted structures.

$CrS_2$ as another important member of TMDs attracts little attention due to the difficult to synthesize bulk and its 2D counterpart. Inspiringly, the 2H, 1T, and 1T' structures coexisting were observed in the monolayer $CrS_2$ prepared via the chemical vapor deposition (CVD) method . Thus, it is much possible that bilayer $CrS_2$ with pure 2H structure is grown via CVD. It is necessary and interesting to study geometric and electronic structures of bilayer $CrS_2$ with different twist angles. We find that no matter what optimization strategy is adopted, the GGA IDs of bilayer $CrS_2$ with any twist angles are larger than that of bilayer $MoS_2$, although atomic radius of Cr is smaller than that of Mo. However, the other five types of IDs for bilayer $CrS_2$ are smaller than corresponding that of bilayer $MoS_2$. All types of IDs (except for GGA) of bilayer $CrS_2$ with zero angle are larger than that of 13.2°, 21.8° and 32.2°, which is the same with bilayer $MoS_2$ case.

Total energy each atom $E_{atom}$ for all twisted structures of TMDs are showed in Fig. 4 and Fig. 5. Although the difference of $E_{atom}$ by A and B strategy are negligible, in the most cases, $E_{atom}$ by A is smaller than that by B. This implies that the A optimization strategy is superior to the B. In the case of the same exchange–correlation functional, the $E_{atom}$ difference between different twist angles structures is also negligible, which can be masked by thermal vibration of atom at ambient temperature. Therefore, from the view of energetics, it is possible to experimentally prepared twisted structures of TMDs. The GGA-vdW-DF and GGA-vdW-DF2 $E_{atom}$ are obviously higher than the other $E_{atom}$. In fact, the comparison between the total energy by different exchange–correlation functional is no point

**B electronic structures**

In this section, we put focus on effect of twist angle on electronic structure and more especially on bandgap. Previous work has reported that bulk $MoS_2$ and its 2D counterparts except monolayer are indirect bandgap semiconductors. Valence band maximum (VBM) and conduction band minimum (CBM) locate at $\Gamma$ and K high-symmetry points in reciprocal space, respectively. Moreover, one local VBM is at K point, so the direct bandgap appear at K point. Thus, there are two obvious photoluminescence (PL) peaks at of 1.6 eV and 1.8 eV corresponding to the indirect and direct bandgaps. It is noted that PL spectra detects the optical bandgap that is different from electronic bandgap also known as the fundamental (or transport) bandgap. As known, the difference comes from strongly bound exciton that origins from strong Coulomb interactions between n- and p-type carriers. Theoretically, the optical bandgap can be defined as electronic bandgap minus the exciton binding energy . Therefore, electronic bandgap is usually larger than the optical bandgap.

Electronic GGA bandgaps of 0° bilayer $MoS_2$ structure by A and by B are 1.58 eV and 1.64 eV that are larger than the other types of bandgaps (showed in Fig. 6). The GGA bandgap by B is larger than the optical bandgap attained by PL, suggesting greater rationality without considering the rationality of the optimized structure. The GGA bandgaps of bilayer $MoS_2$ with twist angles are smaller than that with zero angle. However, the case of the LDA bandgaps is the opposite. This possibly attributes to larger GGA IDs, compared to the LDA IDs. We consider that the larger ID can leads to larger bandgap, which is independent of the choice of exchange-correlation functionals. Namely, for the same structure, the bandgaps attained by GGA and by LDA should have little difference. For vdW correction, the bandgaps of normal structure by A are smaller than that of twisted structures. However, using the B strategy, we obtain the opposite case: the bandgaps of zero angle structure are larger than that of twisted structures. This again implies the present vdW correction method still needs further improvement. The vdW-DF and vdW-DF2 bandgaps are obviously smaller than the other types of bandgaps, because the two vdW corrections belongs to non-local correlation functional that approximately accounts for dispersion interactions.

As shown in Fig. 7, the corresponding bandgaps of bilayer $CrS_2$ are smaller, compared with that of bilayer $MoS_2$. It is understandable that Cr element possess stronger metallic property. The GGA bandgaps of normal bilayer $CrS_2$ are slightly larger than that with twist angles, although the corresponding GGA IDs are obviously larger than IDs of twisted structures. This suggests that the GGA bandgaps are not sensitive to ID of bilayer $CrS_2$. Moreover, in most cases, bandgaps of bilayer $CrS_2$ structures by A and by B are basically consistent with each other.

Sequentially, we focus on the effects of twist angle on the band structures of TMDs. It can be seen from Fig. 8 and Fig. 9 that the GGA band structures of TMDs with twist angles looks clutter because there are more atoms in primitive cells. The twist angles affect the shapes of VB and CB especially near local VBM at K point. The other types of band structures also show twist angle effects, we only plot the GGA-DFT-D2 band structures in the Supplementary Information (see Figs. S1 and S2). Whether using A or B strategy, the twist angles render both VBs and CBs flat, which is consistent with recent theoretical work based on revised tight-binding model. Especially, the shape of VB for bilayer $CrS_2$ with twist angle of 13.2° seems to be a line. It is worth mentioning that the decrease of VB or CB width usually leads to larger bandgap. However, twist angles render narrow band width accompanied by smaller bandgaps. This band feature is useful in some applications. For instance, flat band and small bandgap respectively support high Seebeck coefficient and electrical conductivity that is good for realizing high thermoelectric merit of figure.

As known, spin-orbit coupling (SOC) has an important effect on electronic structures of TMDs. Our GGA-SOC calculations show that there is obvious spin-orbit splitting in the normal and twisted structures (see Fig. S3). The degree of the splitting is sensitive to the twisted angle. For instance, the splitting of $MoS_2$ with 0° and 21.8° reaches up to 155 meV and 165 meV while the splitting in the 13.2° and 32.2° structures is negligible. For CrS$_2$, the largest splitting is only 65 meV in the 21.8° structure. However, there is obvious splitting in the 32.2° structure, which is different from MoS$_2$.

## IV. SUMMARY

In conclusion, we have employed the LDA, pure GGA and GGA with four vdW corrections to explore crystal and electronic structures of TDM twisted structures. Our first-calculations calculations shows pure GGA is better for describing the interlayer distance change caused by effect of twist angle in twisted structures than the other methods, although its results is contradict with the LDA results. The GGA ID and bandgap of twisted structures is smaller than that of normal structure. We find that reduced interlayer distance and twisted angle can lead to flat VB and CB with smaller bandgap, and consider this band feature is possibly valuable in thermoelectric application.

**ACKNOWLEDGMENT:**

This work is supported by the National Natural Science Foundation of China (Grant No. 11804132), the National Natural Science Foundation of China (Grant No. 11847129) and Sichuan Science and Technology Program (Grant No. 2019YJ0336).

**DATA AVAILABILITY**

The data that supports the findings of this study are available within the article and its supplementary material.

**References:**

[1] K. S. Novoselov, A. K. Geim, S. V. Morozov, D. Jiang, Y. Zhang, S. V. Dubonos, I. V. Grigorieva, and A. A. Firsov, Science **306**, 666 (2004).
[2] X. D. Xu, W. Yao, D. Xiao, and T. F. Heinz, Nat. phys. **10**, 343 (2014).
[3] F. F. Zhu, W. J. Chen, Y. Xu, C. L. Gao, D. D. Guan, C. H. Liu, D. Qian, S. C. Zhang, and J. F. Jia, Nat. Mater. **14**, 1020 (2015).

[4] S. Z. Butler *et al.*, Acs Nano **7**, 2898 (2013).
[5] Y. J. Fu *et al.*, Npj Quantum Materials **2**, 52 (2017).
[6] E. W. Lee, R. Kim, J. Ahn, and B. J. Yang, Npj Quantum Materials **5**, 1 (2020).
[7] F. Liu, Y. Zhou, Y. J. Wang, X. Y. Liu, J. Wang, and H. Guo, Npj Quantum Materials **1**, 16004 (2016).
[8] S. Zhou, C. C. Liu, J. J. Zhao, and Y. G. Yao, Npj Quantum Materials **3**, 16 (2018).
[9] T. Zhou, J. Y. Zhang, H. Jiang, I. Zutic, and Z. Q. Yang, Npj Quantum Materials **3**, 39 (2018).
[10] L. N. Du *et al.*, 2d Materials **6**, 025014 (2019).
[11] Y. F. Wu *et al.*, Nat. Commu. **7**, 10270 (2016).
[12] M. R. Brems and M. Willatzen, New J. Phys. **21**, 093030 (2019).
[13] R. M. Zhao, T. X. Wang, M. Y. Zhao, C. X. Xia, Y. P. An, S. Y. Wei, and X. Q. Dai, Appl. Surf. Sci. **491**, 128 (2019).
[14] Z. D. Song, Z. J. Wang, W. J. Shi, G. Li, C. Fang, and B. A. Bernevig, Phys. Rev. Lett. **123**, 036401 (2019).
[15] G. Tarnopolsky, A. J. Kruchkov, and A. Vishwanath, Phys. Rev. Lett. **122**, 106405 (2019).
[16] S. X. Huang, X. Ling, L. B. Liang, J. Kong, H. Terrones, V. Meunier, and M. S. Dresselhaus, Nano Lett. **14**, 5500 (2014).
[17] P. C. Yeh *et al.*, Nano Lett. **16**, 953 (2016).
[18] S. Horzum, H. Sahin, S. Cahangirov, P. Cudazzo, A. Rubio, T. Serin, and F. M. Peeters, Phys. Rev. B **87**, 125415 (2013).
[19] H. J. Conley, B. Wang, J. I. Ziegler, R. F. Haglund, S. T. Pantelides, and K. I. Bolotin, Nano Lett. **13**, 3626 (2013).
[20] A. Ramasubramaniam, D. Naveh, and E. Towe, Phys. Rev. B **84**, 205325 (2011).
[21] R. Bistritzer and A. H. MacDonald, Proc. Natl. Acad. Sci. U.S.A. **108**, 12233 (2011).
[22] X. Z. Zhang, R. Y. Zhang, Y. Zhang, T. Jiang, C. Y. Deng, X. A. Zhang, and S. Q. Qin, Opt. Mater. **94**, 213 (2019).
[23] D. A. Ruiz-Tijerina and V. I. Fal'ko, Phys. Rev. B **99**, 125424 (2019).
[24] M. H. Naik and M. Jain, Phys. Rev. Lett. **121**, 266401 (2018).
[25] S. Z. Zhu and H. T. Johnson, Nanoscale **10**, 20689 (2018).
[26] Y. P. Zhao, C. W. Liao, and G. Ouyang, J. Phys. D. Appl. Phys. **51**, 185101 (2018).
[27] F. C. Wu, T. Lovorn, E. Tutuc, I. Martin, and A. H. MacDonald, Phys. Rev. Lett. **122**, 086402 (2019).
[28] Z. Bi, N. F. Q. Yuan, and L. Fu, Phys. Rev. B **100**, 035448 (2019).
[29] I. Maity, M. H. Naik, P. K. Maiti, H. R. Krishnamurthy, and M. Jain, Physical Review Research **2**, 013335 (2020).
[30] C. X. Liu, Phys. Rev. Lett. **118**, 087001 (2017).
[31] Y. H. Tan, F. W. Chen, and A. W. Ghosh, Appl. Phys. Lett. **109**, 32 (2016).
[32] B. X. Cao and T. S. Li, Journal of Physical Chemistry C **119**, 1247 (2015).
[33] C. H. Jin *et al.*, Nature **567**, 76 (2019).

[34] Z. Z. Jiang, W. D. Zhou, A. J. Hong, M. M. Guo, X. F. Luo, and C. L. Yuan, Acs Energy Letters **4**, 2830 (2019).

[35] A. J. Hong, Appl. Surf. Sci. **473**, 6 (2019).

[36] F. Sun, A. Hong, W. Zhou, C. Yuan, and W. Zhang, Materials Today Communications **25**, 101707 (2020).

[37] G. Kresse and J. Furthmüller, Phys. Rev. B **54**, 11169 (1996).

[38] G. Kresse and J. Furthmüller, Comp. Mater. Sci. **6**, 15 (1996).

[39] J. P. Perdew and Y. Wang, Phys. Rev. B **45**, 13244 (1992).

[40] J. P. Perdew, K. Burke, and M. Ernzerhof, Phys. Rev. Lett. **77**, 3865 (1996).

[41] G. T. de Laissardiere, D. Mayou, and L. Magaud, Nano Lett. **10**, 804 (2010).

[42] X. Y. Zhao, L. Y. Li, and M. W. Zhao, J. Phys-condens. Mat. **26**, 095002 (2014).

[43] J. M. B. L. dos Santos, N. M. R. Peres, and A. H. Castro, Phys. Rev. Lett. **99**, 256802 (2007).

[44] J. Xiao, M. Q. Long, X. M. Li, Q. T. Zhang, H. Xu, and K. S. Chan, J. Phys-condens. Mat. **26**, 405302 (2014).

[45] S. F. Wu *et al.*, Nat. phys. **9**, 149 (2013).

[46] M. R. Habib, S. P. Wang, W. J. Wang, H. Xiao, S. M. Obaidulla, A. Gayen, Y. Khan, H. Z. Chen, and M. S. Xu, Nanoscale **11**, 20123 (2019).

[47] T. Cheiwchanchamnangij and W. R. L. Lambrecht, Phys. Rev. B **85**, 205302 (2012).

[48] S. Venkateswarlu, A. Honecker, and G. T. de Laissardiere, Phys. Rev. B **102**, 081103 (2020).

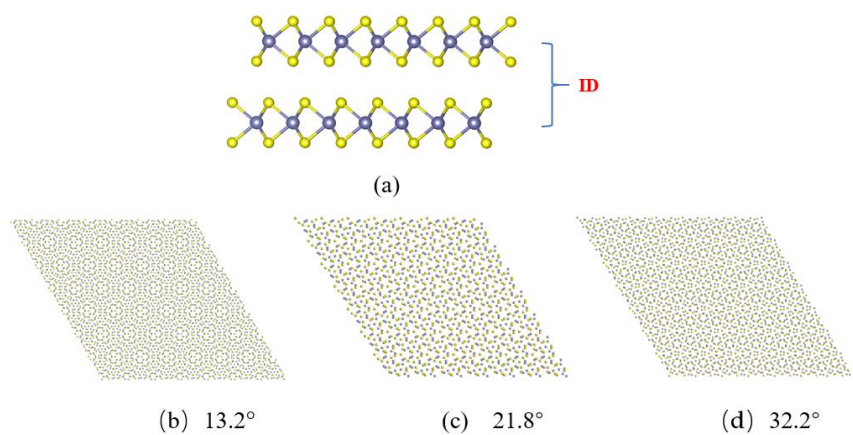

Fig. 1 (Color online) Representation of interlayer distance for bilayer TMD (a), twisted structures with three twist angles 13.2°, 21.8° and 32.2°: (b)-(e).

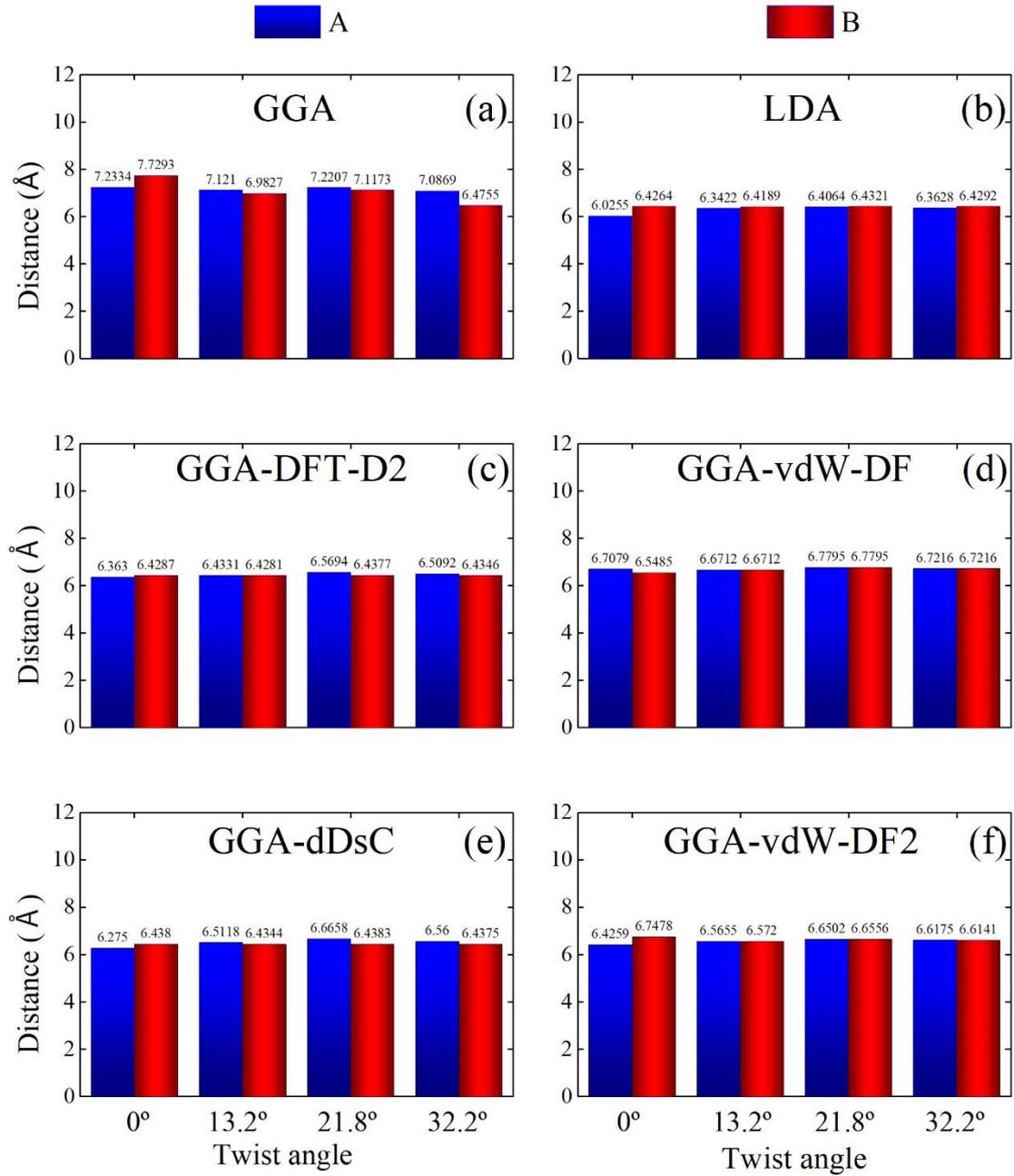

Fig. 2. (Color online) Interlayer distances of bilayer MoS$_2$ by A and by B using six exchange correlation functionals GGA (a), LDA (b), GGA-DFT-D2 (c), GGA-vdW-DF (d), GGA-dDsC (e) and GGA-vdW-DF2 (f).

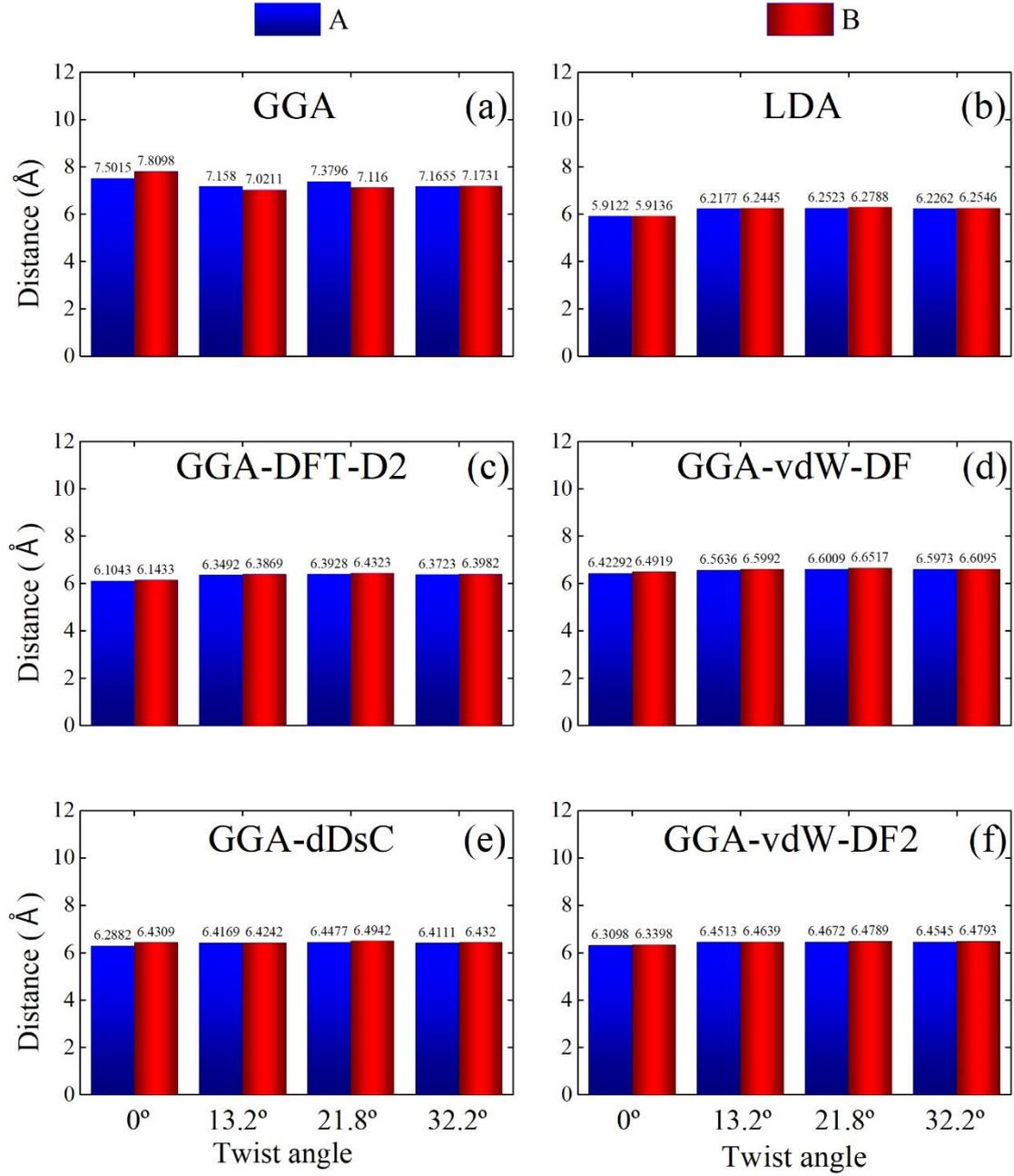

Fig. 3. (Color online) Interlayer distances of bilayer CrS$_2$ by A and by B using six exchange correlation functionals GGA (a), LDA (b), GGA-DFT-D2 (c), GGA-vdW-DF (d), GGA-dDsC (e) and GGA-vdW-DF2 (f).

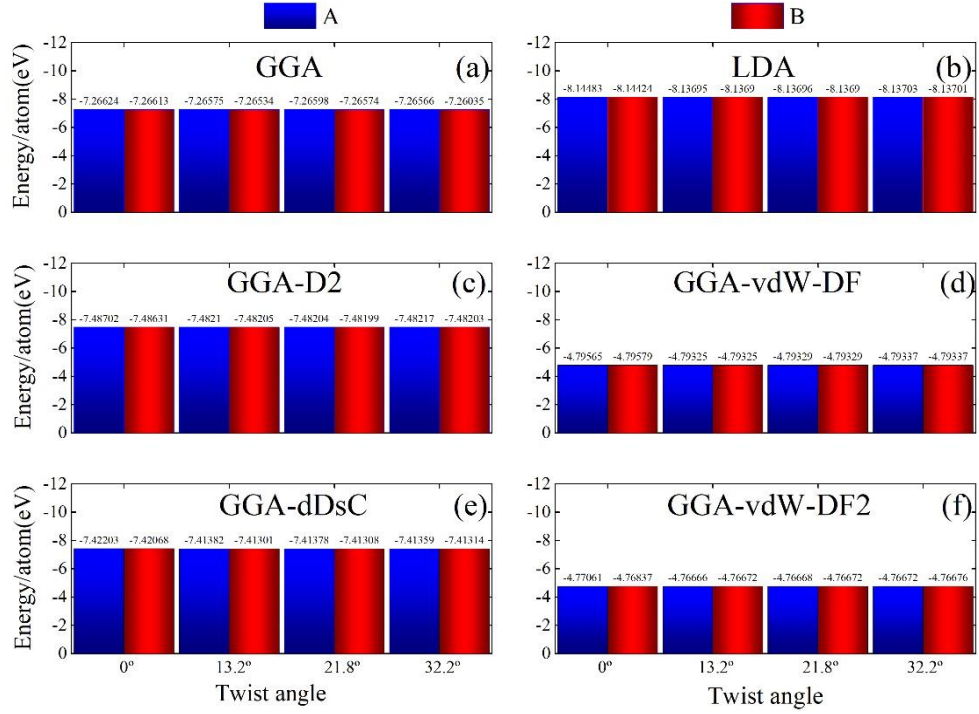

Fig. 4. (Color online) Total energies each atom of bilayer MoS$_2$ by A and by B using six exchange correlation functionals GGA (a), LDA (b), GGA-DFT-D2 (c), GGA-vdW-DF (d), GGA-dDsC (e) and GGA-vdW-DF2 (f).

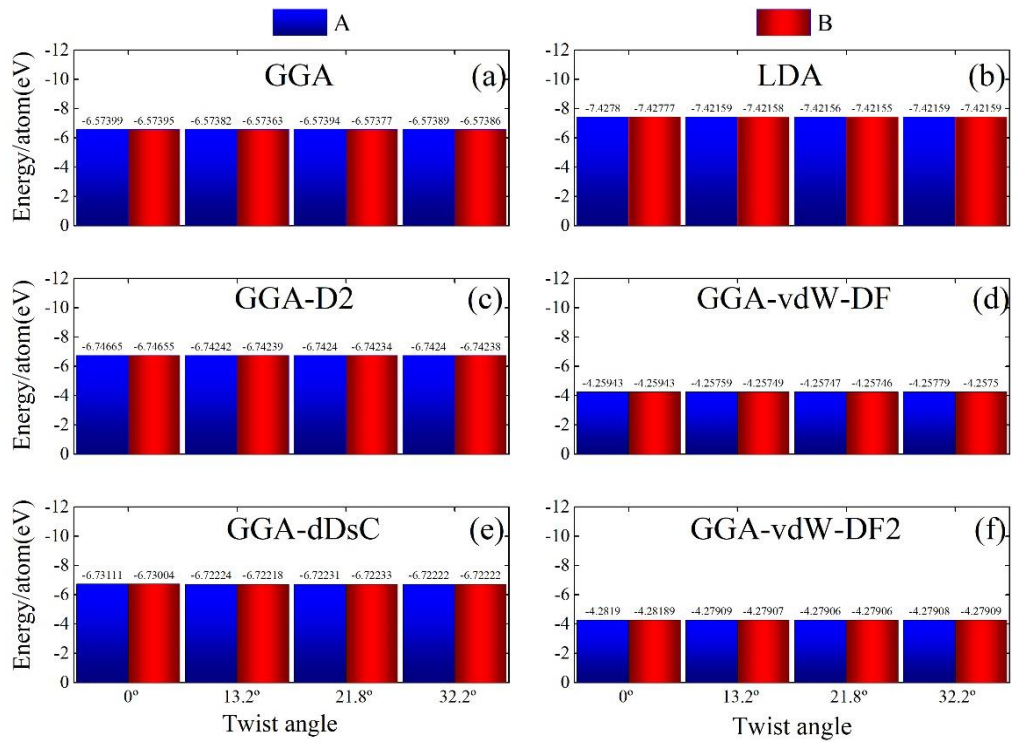

Fig. 5 (Color online) Total energies each atom of bilayer CrS$_2$ by A and by B using six exchange correlation functionals GGA (a), LDA (b), GGA-DFT-D2 (c), GGA-vdW-DF (d), GGA-dDsC (e) and GGA-vdW-DF2 (f).

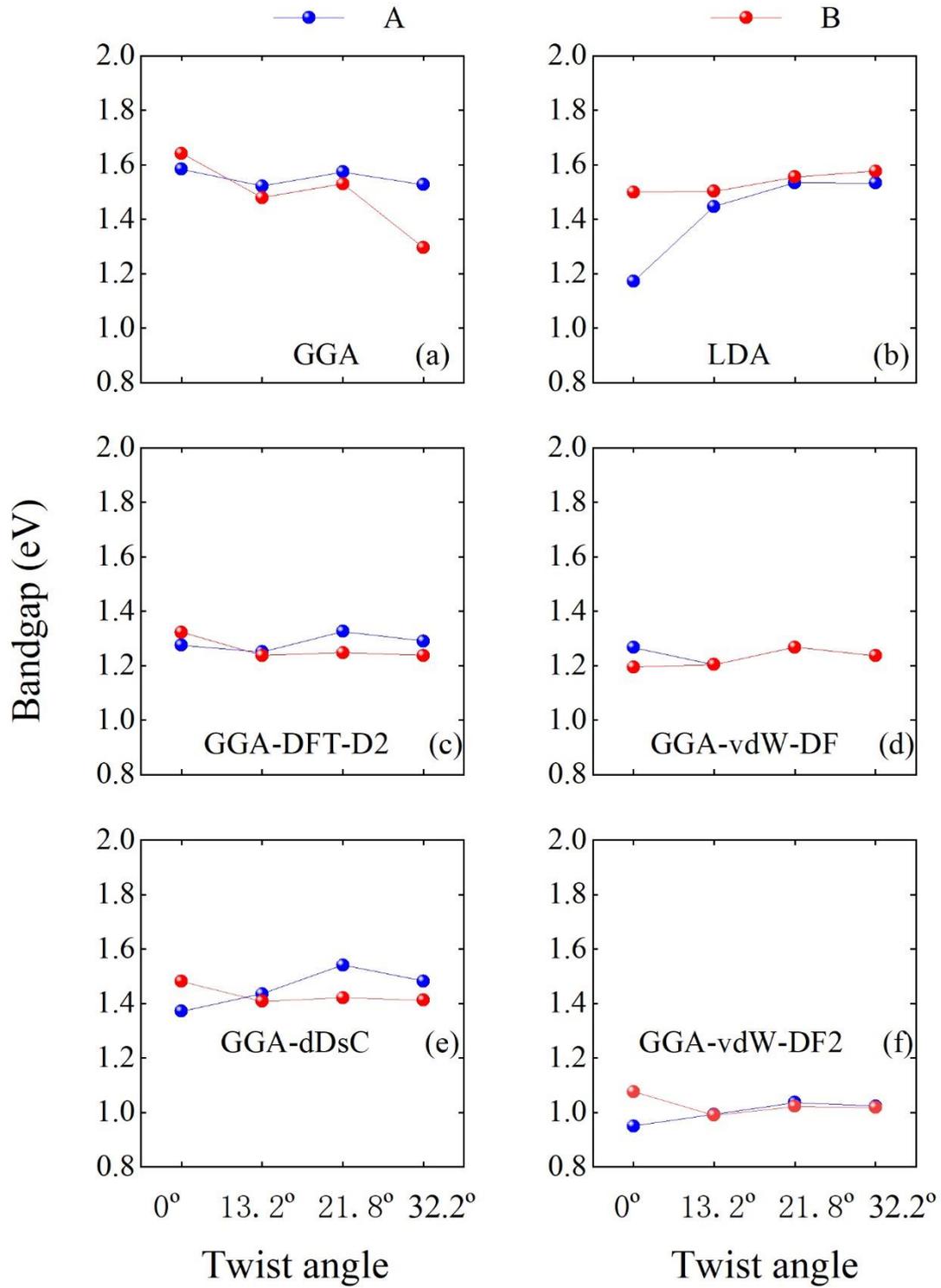

Fig. 6 (Color online) Bandgaps of bilayer $MoS_2$ by A and by B using six exchange correlation functionals GGA (a), LDA (b), GGA-DFT-D2 (c), GGA-vdW-DF (d), GGA-dDsC (e) and GGA-vdW-DF2 (f).

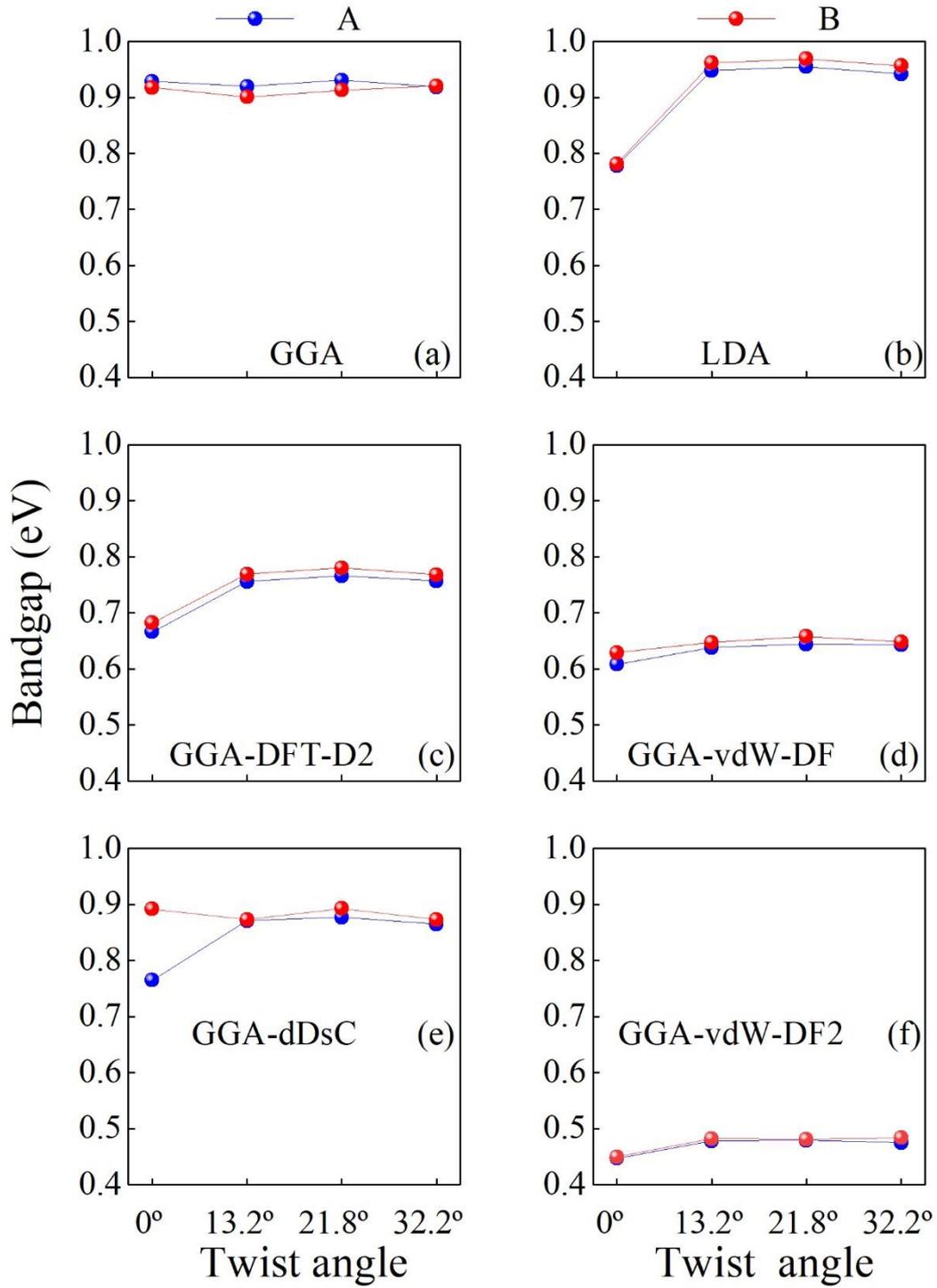

Fig. 7 (Color online) Bandgaps of bilayer $CrS_2$ by A and by B using six exchange correlation functionals GGA (a), LDA (b), GGA-DFT-D2 (c), GGA-vdW-DF (d), GGA-dDsC (e) and GGA-vdW-DF2 (f).

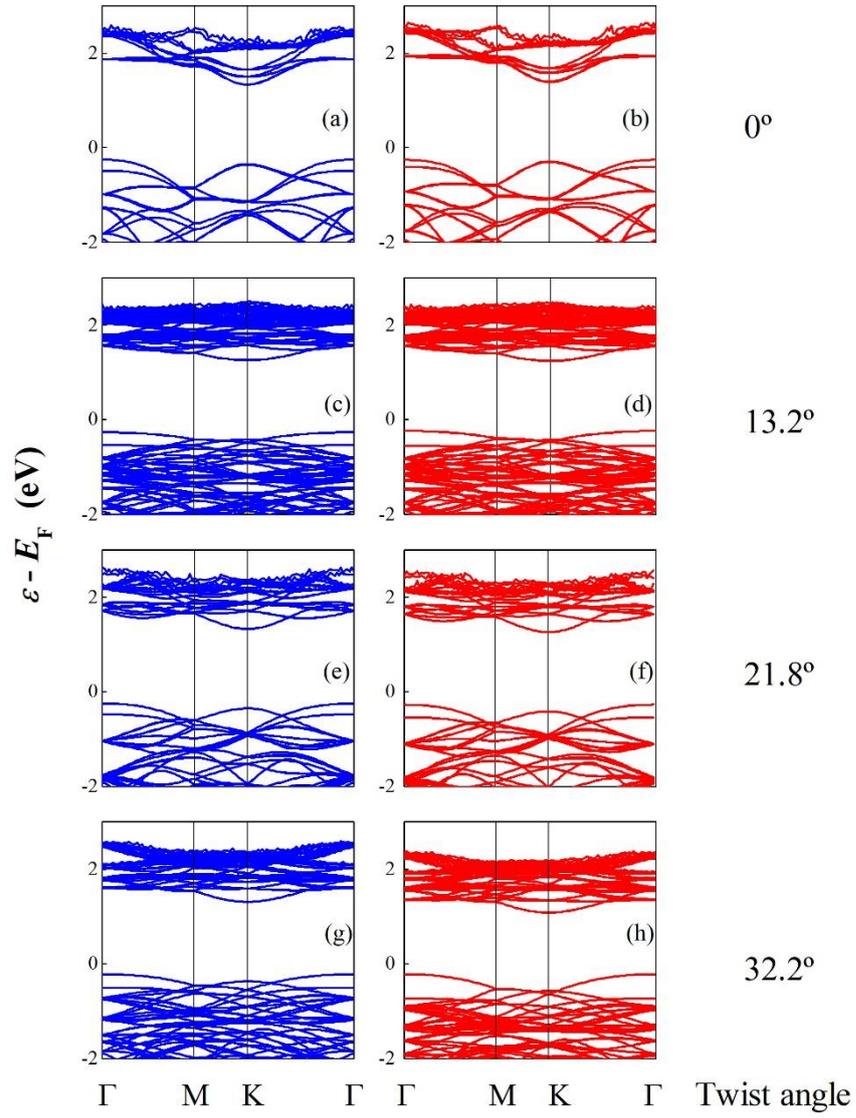

Fig. 8 (Color online) GGA Band structures of bilayer MoS$_2$ with twist angles 0° (top row), 13.2° (the second row), 21.8° (the third row) and 32.2° (bottom row). The left and right columns correspond to band structures of bilayer by A and by B, respectively.

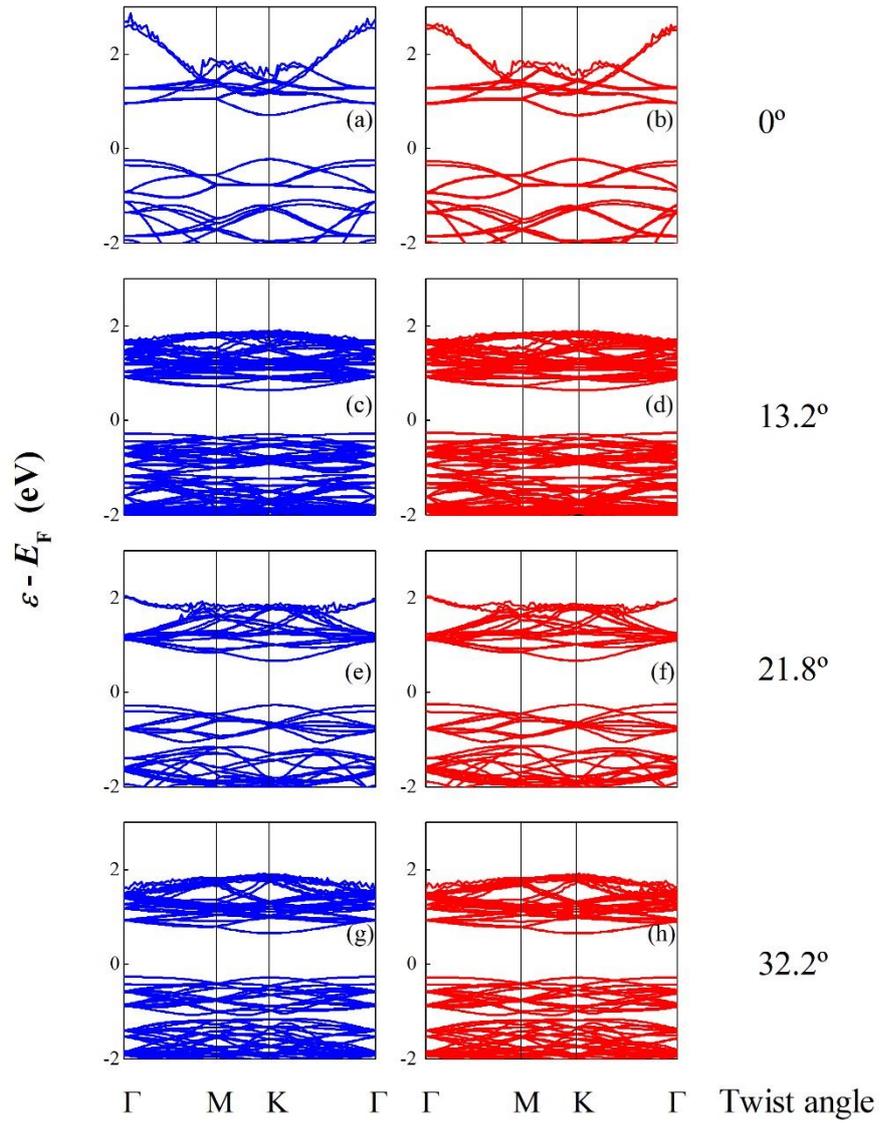

Fig. 9 (Color online) GGA Band structures of bilayer MoS$_2$ with twist angles 0° (top row), 13.2° (the second row), 21.8° (the third row) and 32.2° (bottom row). The left and right columns correspond to band structures of bilayer by A and by B, respectively.